\documentclass{article}
\usepackage{arxiv}

\usepackage[utf8]{inputenc} 
\usepackage[T1]{fontenc}    
\usepackage{hyperref}       
\usepackage{url}            
\usepackage{booktabs}       
\usepackage{amsfonts}       
\usepackage{nicefrac}       
\usepackage{microtype}      
\usepackage{lipsum}
\usepackage{graphicx}
\usepackage{natbib}
\usepackage{dsfont}
\usepackage{amsmath, amssymb}
\usepackage{algorithm, algorithmicx, algpseudocode}
\usepackage{multirow}

\title{A Non-Stationary Spatio-Temporal Covariance Model with Dynamic Advection Effects for Rainfall Data}

\author{
 Pedro Nasevicius Ramos \\
  Department of Statistics -- IMECC\\
  University of Campinas\\
  S\~{a}o Paulo, Brazil \\
  \texttt{ra185812@dac.unicamp.br} \\
   \And
 Guilherme Ludwig \\
  Department of Statistics -- IMECC\\
  University of Campinas\\
  S\~{a}o Paulo, Brazil \\
  \texttt{gvludwig@unicamp.br} \\
}

\begin{document}

\maketitle

\abstract{We propose a non-stationary model constructed using a mixture of spatio-temporal covariance models with advection effects; namely, models that have larger covariance values along an orientation vector in the spatio-temporal index set, that simulate wind direction and cloud movement. We show that a mixture of such models can allow for wind direction change in data during (estimated) time intervals, unlike traditional models that use rigid advection effects. We construct a MCMC procedure for Bayesian estimation, and illustrate the method with the analysis of a severe rainfall event from the southeastern region of Brazil.}


\section{Introduction}\label{sec1}

Rain is a natural phenomenon that has been studied since antiquity, with an immense impact on agriculture and the life of a biome. Rainfall is the primary source of water replenishment for most hydrological systems. Conversely, both drought conditions and excessive rainfall (storms) have destructive impacts. Models for rainfall, either for short and long term forecasting, both deterministic and stochastic, are of interest \citep[see, for example,][]{durran2010numerical}. While widely used, deterministic numerical models of rainfall often require a high degree of computational complexity and detailed knowledge of initial conditions and the physical processes involved. 

Spatiotemporal stochastic models offer an alternative approach, enabling the analysis of spatiotemporal dependence among observations and the generation of probabilistic forecasts that reflect associated uncertainties. Furthermore, stochastic simulation tends to be relatively less complex than deterministic simulation, yielding representative simulations with fewer inputs. Pioneering statistical rainfall models include \citet{gupta1987taylor}, \citet{rodriguez1987some}, \citet{cox1988spatial}, and \citet{wilks1998multisite}. The fusion of statistical and deterministic models in the context of rainfall is also examined by \citet{onof1994note}, while general statistical methodology for merging deterministic and stochastic models is addressed by \citet{poole2000inference}. Among other references on stochastic modeling—including the advantages of simulation—the review article by \citet{northrop2024stochastic} (and the references cited therein) is recommended, as is \citet{salvana2020nonstationary}.

Separable spatio-temporal covariance models are a restrictive class of covariance functions, defined as the product of a strictly spatial covariance function and a strictly temporal one. While separability allows for easier computation, especially when using likelihood-based approaches \citet[p.310]{cressie2011statistics}, it is considered an unrealistic assumption, especially in fields such as atmospheric sciences and epidemiology. Non-separable flexible models and methods to assess non-separability of data have been studied in recent years \citep{o1998markov,gneiting2006geostatistical,bourotte2016flexible,iaco2019isotropy}.

For rainfall data, short term rainfall phenomena may satisfy Taylor's hypothesis \citep{taylor1938spectrum}, a less restrictive type of symmetry, for which the correlation decay rate between two distinct locations, as we consider locations further apart, is proportional to a correlation decay in time between measurements at the same location. Taylor's hypothesis does not always hold, particularly in South American forests \citep{poveda2005testing}, but it is a reasonable assumption when studying rainfall during short periods for which wind changes have a strong effect in cloud movement \citep{northrop2024stochastic}.

Covariance models with advection effects have been introduced by \citet{gupta1987taylor} and \citet{cox1988spatial}. Recently, \citet{salvana2020nonstationary} extended such models to the multivariate case with a Lagrangian framework, including scenarios with multiple advection effects in \citet{salvana2023spatio}. Covariance models with advection effects can also be used in other contexts, such as the work on solar irradiance by \citet{kunke2020nonstationary}, among others.

In this work, we propose a new spatio-temporal covariance model that incorporates many advection effects, and is suitable to precipitation data where rainfall general direction changes over time. The proposed model generalizes the class of Lagrangian spatio-temporal covariance functions first introduced by \citet{gupta1987taylor} and \citet{cox1988spatial} by allowing the advection vector, as well as the dependence and variance parameters, to vary dynamically over time. Furthermore, the model is capable of estimating change points in the spatio-temporal pattern, which is relevant for identifying local climate transitions.

\section{A Review of Spatio-temporal models with advection}\label{sec2}

Let $t \in \mathcal{T} \subset \mathbb{R}$ and $\mathbf{s} \in \mathcal{S} \subset \mathbb{R}^2.$ Define space-time covariance function as \[C(\mathbf{s},\mathbf{s}^\prime; t, t^\prime) = \text{Cov}\left(Y(\mathbf{s},t), Y(\mathbf{s}^\prime,t^\prime) \right).\] 
A simple model of space-time covariances is the \textbf{separable model}, where \[C(\mathbf{s},\mathbf{s}^\prime, t, t^\prime) = C_S(\mathbf{s},\mathbf{s}^\prime)C_T(t, t^\prime).\] This formulation has computational advantages: A separable model allows us to choose known spatial and temporal covariance functions. Moreover, separability also allows for an easier computational implementation, in particular with regards to likelihood-based approaches, since for a dataset with identical resolution and equal number of temporal observations across sites, one would have $\boldsymbol\Sigma_T \otimes \boldsymbol\Sigma_S$ covariance matrix. However, separability implies complete symmetry (CS): that for all \(\mathbf{s},\mathbf{s}^\prime \in \mathbb{R}^d\) and \(t,t^\prime \in \mathbb{R},\) we have \[\text{Cov}\left(Y(\mathbf{s},t), Y(\mathbf{s}^\prime,t^\prime) \right) = \text{Cov}\left(Y(\mathbf{s},t^\prime), Y(\mathbf{s}^\prime,t) \right).\] Complete symmetry is limiting: assume, for example, that clouds travel 25km/h towards east. If we model rainfall using a stochastic process, we expect that the correlation of rainfall between Paraty (23.224 S; 44.727 W) at noon and Rio de Janeiro (22.988 S; 43.191 W) around 5pm (approximately 250 km apart, Paraty is west of Rio) to be higher than the correlation between Rio de Janeiro around 5pm and Paraty at noon. However a CS model postulates they are the same.

Non-separable stationary and non-stationary spatio-temporal covariance functions can be constructed based on \citet{cressie1999classes}, \citet{gneiting2002nonseparable}, and others. In this work, we focus on the class of functions that satisfy Taylor's hypothesis \citep{taylor1938spectrum}: that there exists a vector \(\mathbf{V} \in \mathbb{R}^d\) such that \[C(\mathbf{0}; u) = C(\mathbf{V}u; 0).\] The units of \(\mathbf{V}\) are the ratio between units of \(\mathbf{h}\) and \(u,\) space / time, thus \(\mathbf{V}\) can be interpreted as a speed vector. The advection class of spatio-temporal covariance functions was introduced in \citet{gupta1987taylor}, and \citet{cox1988spatial} extended it; both cases consider univariate Gaussian random fields \citep[the multivariate case was studied in ][]{salvana2020nonstationary,salvana2023spatio}. Consider a random field with finite variance \[Y(\mathbf{s},t) = \tilde{Y}(\mathbf{s} - \mathbf{V} t), \qquad (\mathbf{s},t) \in \mathbb{R}^d \times \mathbb{R}, \mathbf{V} \in \mathbb{R}^d, d \geq 1,\] where \(\tilde{Y}(\mathbf{s} )\) is a purely spatial random field, and \(\mathbf{V}\) is the advection vector that controls speed and transport direction for \(\tilde{Y}(\mathbf{s} )\). Its spatio-temporal covariance function is given by \[C(\mathbf{h},u) = C_S(\mathbf{h} - \mathbf{V} u), \qquad \mathbf{h} = \mathbf{s}^\prime-\mathbf{s}, u= t^\prime-t,\] where \(C_S\) is a purely spatial covariance function.

We can treat \(\mathbf{V}\) as a constant vector, and thus the associated random processes are called \textit{frozen fields} \citep{gupta1987taylor}. If we treat \(\mathbf{V}\) as a random variable, the associated random processes are called \textit{non-frozen fields} \citep{cox1988spatial}. Treating $\mathbf{V}$ as a random variable allow the transported random field to show some diffusion in time, unlike the frozen fields which are rigidly transported along $\mathbf{V}.$ For example, an exponential covariance model using \citet{gupta1987taylor} formulation would be \begin{equation}
\label{eq:cov_frozen}
    C(\mathbf{h},u) = \varphi(\mathbf{h}-\mathbf{V} u) =  \sigma^2 \exp\left \{-\frac{\|\mathbf{h}-\mathbf{V} u\|}{\theta} \right\}, 
\end{equation} with paramters $\sigma^2 > 0,$ for spatial variance, and $\theta > 0$ for spatial dependence. On the other hand, for a \textit{non-frozen field} case of \citet{cox1988spatial}, the advection vector is sampled from \(\mathbf{V} \sim N(\boldsymbol\mu_{\mathbf{V}} ,\boldsymbol\Sigma_{\mathbf{V}} )\), where \(\boldsymbol\mu)\) controls the expected direction of the advection vector, and $\boldsymbol\Sigma_{\mathbf{V}}$ controls the process' diffusion. The expected spatio-temporal covariance function is thus \begin{equation}
\label{eq:cov_nonfrozen}
    C(\mathbf{h},u)= \frac{1}{\sqrt{|\boldsymbol{I}_{2\times 2} +  \boldsymbol\Sigma_{\mathbf{V}}u^2}|} \varphi \left (  \left[ (\mathbf{h}-\boldsymbol\mu u)^T (\boldsymbol{I}_{2\times 2} + \boldsymbol\Sigma_{\mathbf{V}}u^2)^{-1} (\mathbf{h}-\boldsymbol\mu u) \right]^{1/2} \right),
\end{equation} where \(\varphi(\cdot)\) is a spatial covariance function, such as the exponential in \ref{eq:cov_frozen}), with parameters \(\sigma^2>0\) and \(\theta>0\). The covariance function in \eqref{eq:cov_nonfrozen} can lead to dense covariance matrices, however, and be difficult to estimate in general.

\section{A model with multiple advection effects}\label{sec3}

We propose a flexible model that allows for advection from multiple sources. This model adapts the data fusion framework of \citet{chu2014semiparametric} and \citet{ludwig2017static}. For a fixed \(L,\) let \begin{equation*} 
        C(\textbf{h},u) = \sum_{k=1}^L\lambda_k c_k(\textbf{h}- \textbf{V}_k u)\phi_k(t)\phi_k(t'),
        \label{eq:modelCS}
\end{equation*} where \(\textbf{h} = \textbf{s}-\textbf{s}'\), \(u=t-t'\), and for \(k=1,2,\ldots, L,\) \(\lambda_k >0,\) and $c_k(\cdot)=\text{exp} \left\{- \left \| \cdot \right \|/ \theta_k \right\}.$ We assume $t \in [0,T],$ and that the functions \(\phi_k\) are step functions of the form $\phi_k(t) = \mathbf{1}\{t \leq \tau_k\},$ with unknown \(\tau_k,\) for $k=1,2,\ldots,L-1,$ and $\phi_L(t) = \mathbf{1}\{t > \tau_{L-1}\}.$ This choice of $\phi_k$ leads to a direct Gibbs sampler, but other choices may be used, such as spline functions \citep[see][]{ludwig2017static}.

Let $\mathbf{s}_1, \ldots, \textbf{s}_n$ be the $n$ spatial sites, each with complete observations indexed by $0 = t_1 < \cdots < t_m = T.$ Denote by $\boldsymbol\zeta$ the array of parameters $\boldsymbol\zeta = (L,\boldsymbol\lambda, \boldsymbol\theta, \mathbf{V}_1, \ldots, \mathbf{V}_L, \boldsymbol\tau).$ Let \(\mathbf{Z} = \{Z(\textbf{s}_1,t_1), \dots, Z(\textbf{s}_n,t_m)\}^T\) and \(\boldsymbol\Sigma = \boldsymbol\Sigma(L,\boldsymbol\lambda, \boldsymbol\theta, \mathbf{V}_1, \ldots, \mathbf{V}_L, \boldsymbol\tau) = \boldsymbol\Sigma(\boldsymbol\zeta) = \text{Cov}(\mathbf{Z}).\)
Let \(\mathbf{Z}|\boldsymbol\Sigma \sim N(\mathbf{0},\boldsymbol\Sigma).\) We considered a Bayesian framework to estimate the parameters, and implemented the Gibbs sampler with Metropolis steps. The following priors are used: \begin{align*}
    \lambda_k|a_k,b_k \sim \text{Inv-Gamma}(a_k,b_k), \qquad & \theta_k|g_k,\omega_k \sim \text{Gamma}(g_k,\omega_k),\\
    \mathbf{V}_k|\mu_k, \sigma_k^2 \sim N(\mu_k,\sigma^2_k\mathbf{I}_2 ), \qquad & \boldsymbol\tau \sim \mathds{1}\{0 \leq \tau_1 < \cdots < \tau_{L-1} \leq T\}. \end{align*} Observe the joint posterior distribution of parameters is given by 
\begin{equation}
\begin{aligned}
    \pi(\boldsymbol{\theta},\boldsymbol{\lambda},\boldsymbol{V}, \boldsymbol\tau|\mathbf{Z}, \text{others}) \propto &  
   \left[\prod_{k=1}^L \pi(\theta_k|\omega_k) \pi(\lambda_k|a_k,b_k) \pi(V_k|\sigma^2_k,\mu_k) \right] \times \pi(\boldsymbol{\tau})\times \mathcal{L}(\mathbf{Z}|\boldsymbol\zeta) \\
    \propto & 
    \prod_{k=1}^L \underset{\text{priors for }\theta_k }{\underbrace{
    e^{-\omega_k \theta_i} \mathds{1}_{\{ \theta_k>0\}}
    }} \underset{\text{priors for }\lambda_k }{\underbrace{
    \left( \frac{1}{\lambda_k} \right)^{a_k+1}\exp\left\{-\frac{b_k}{\lambda_k}\right\} \mathds{1}_{\{\lambda_k > 0\}}
    }} \\
    & \times \underset{\text{priors for }  V_k }{\underbrace{
     |\sigma^2_k\textbf{I}_2|^{-1/2} \exp\left\{  -\frac{1}{2\sigma^2_k} (V_k - \mu_k)^T(V_k - \mu_k) \right\} 
    }} \\ 
    & \times \underset{\text{priors for }\boldsymbol\tau }{\underbrace{
    \mathds{1}\{0 \leq \tau_1 < \cdots < \tau_{L-1} < T\}
    }}
    \underset{\text{likelihood}}{\underbrace{
    |\Sigma(\boldsymbol\zeta)|^{-1/2} \exp\left\{  -\frac{1}{2} \mathbf{Z}^T\Sigma(\boldsymbol\zeta)^{-1}\mathbf{Z} \right\} 
    }}
\end{aligned}
\label{eq:full}
\end{equation}

The full conditionals based on Equation \eqref{eq:full} and the Gibbs sampler is shown in Algorithm \ref{alg:gibbs}. We remark that the full conditionals of \(\lambda_k|\text{others},\) \(\theta_k|\text{others},\) \(\mathbf{V}_k|\text{others}\) and \(\boldsymbol\tau|\text{others}\) require Metropolis steps. We used log-normal proposals for the positive parameters $\lambda_k, \theta_k,$ as the Hastings correction is straightforward in those cases, and multivariate normal proposals for $\mathbf{V}_k.$

\begin{algorithm}[!htp]
\caption{Gibbs sampler for the multiple advection model}\label{alg:gibbs}
\begin{algorithmic}
\Require{$\text{Initialize } \lambda^{(1)}_k,\theta^{(1)}_k,\mathbf{V}^{(1)}_k,\boldsymbol{\mu}_k,\sigma^{2}_k,\tau_k, \text{ for } k=1,\dots,L;$ \text{ initialize } $\Sigma^{(1)},\boldsymbol\upsilon_1$}
\For {$\ell$=1,\dots,M} 
   \For {k=1,\dots,L} 
   \State  $\lambda^\ast \sim LN(\lambda_k^{(\ell-1)},\boldsymbol\upsilon_\ell^{(\lambda_k)})$  \Comment{$\boldsymbol\upsilon_\ell$ is an adaptive proposal size; LN = log-normal proposals}
   \State $\text{U} \sim U(0,1)$
   \State $\Sigma^\ast \gets \Sigma(\lambda_1^{(\ell)},\theta_1^{(\ell)},\textbf{V}_1^{(\ell)}, \tau_1^{(\ell-1)}, \dots, \lambda^\ast,\theta_k^{(\ell-1)},\textbf{V}_k^{(\ell-1)}, \tau_k^{(\ell-1)}, \dots, \tau_{L-1}^{(\ell-1)}, \lambda_L^{(\ell-1)},\theta_L^{(\ell-1)},\textbf{V}_L^{(\ell-1)})$
   \State $\alpha \gets \left (\frac{\lambda_k^{(\ell-1)}}{\lambda_k^\ast}\right)^{a_k}  \exp \left \{-\frac{b_k}{\lambda_k^\ast} + \frac{b_k}{\lambda_k^{(\ell-1)}}\right\} \frac{|\Sigma^{(\ell-1)}|^{1/2}}{|\Sigma^{\ast}|^{1/2}} \exp\left\{- \frac{1}{2} \mathbf{Z}^T(\Sigma^{\ast^{-1}}\!-\Sigma^{(\ell-1)^{-1}})\mathbf{Z} \right\}$
   \If{$\text{U} \leq \min(1,\alpha)$}
        \State $\lambda^{(\ell)}_k \gets \lambda^\ast$ and $\boldsymbol\Sigma^{(\ell)} \gets  \boldsymbol\Sigma^{\ast}$
   \Else    
        \State $\lambda^{(\ell)}_k \gets \lambda_k^{(\ell-1)}$
   \EndIf
   \State $\theta^\ast \sim LN(\theta_k^{(\ell-1)},\boldsymbol\upsilon_\ell^{(\theta_k)})$ \Comment{$\boldsymbol\upsilon_\ell$ is an adaptive proposal size; LN = log-normal proposals}
   \State $\text{U} \sim U(0,1)$
   \State $\Sigma^\ast \gets \Sigma(\lambda_1^{(\ell)},\theta_1^{(\ell)},\textbf{V}_1^{(\ell)}, \tau_1^{(\ell-1)}, \dots, \lambda_k^{(\ell-1)},\theta^\ast,\textbf{V}_k^{(\ell-1)}, \tau_k^{(\ell-1)}, \dots, \tau_{L-1}^{(\ell-1)},  \lambda_L^{(\ell-1)},\theta_L^{(\ell-1)},\textbf{V}_L^{(\ell-1)})$
   \State $\alpha \gets \exp\left\{ w_k(\theta_k^{(\ell-1)} - \theta^\ast) \right\} \left(\frac{\theta^\ast}{\theta_k^{(\ell-1)}}\right)^{g_k} \frac{|\Sigma^{(\ell-1)}|^{1/2}}{|\Sigma^{\ast}|^{1/2}} \exp\left\{ - \frac{1}{2} \mathbf{Z}^T (\Sigma^{\ast^{-1}} - \Sigma^{(\ell-1)^{-1}})\mathbf{Z} \right\}$ 
   \If{ $\text{U} \leq \min(1,\alpha)$ }
        \State $\theta^{(\ell)}_k \gets \theta^\ast$ and $\boldsymbol\Sigma^{(\ell)} \gets  \boldsymbol\Sigma^{\ast}$
   \Else
        \State $\theta^{(\ell)}_k \gets \theta_k^{(\ell-1)}$
   \EndIf
   \State $\textbf{V}^{\ast} \sim N(\textbf{V}_k^{(\ell-1)},\boldsymbol\upsilon_\ell^{(\textbf{V}_k)} \textbf{I}_{2 \times 2})$ \Comment{$\boldsymbol\upsilon_\ell$ is an adaptive proposal variance}
   \State $\text{U} \sim U(0,1)$
   \State $\Sigma^\ast \gets \Sigma(\lambda_1^{(\ell)},\theta_1^{(\ell)},\textbf{V}_1^{(\ell)}, \tau_1^{(\ell-1)}, \dots, \lambda_k^{(\ell-1)},\theta_k^{(\ell-1)},\textbf{V}^\ast, \tau_k^{(\ell-1)}, \dots, \tau_{L-1}^{(\ell-1)},  \lambda_L^{(\ell-1)},\theta_L^{(\ell-1)},\textbf{V}_L^{(\ell-1)})$
   \State $\displaystyle\begin{aligned}
       \alpha \gets & \exp\left\{ \frac{1}{2\sigma_k^{2^{(\ell-1)}}} \left(-\textbf{V}^{\ast^T} \textbf{V}^{\ast} +2\textbf{V}^{\ast^T}  \boldsymbol{\mu}_k + \textbf{V}^{(\ell-1)^T}_k  \textbf{V}^{(\ell-1)}_k - 2\textbf{V}^{(\ell-1)^T}_k  \boldsymbol{\mu}_k\right)   \right\}  \\ & \times 
       \frac{|\Sigma^{(\ell)}|^{1/2}}{|\Sigma^{\ast}|^{1/2}} \exp\left\{ - \frac{1}{2} \mathbf{Z}^T(\Sigma^{\ast^{-1}}-\Sigma^{(\ell)^{-1}})\mathbf{Z} \right\}
   \end{aligned}$
   \If{$\text{U} \leq \min(1,\alpha)$}
        \State $\textbf{V}^{(\ell)}_k \gets \textbf{V}^\ast$ and $\boldsymbol\Sigma^{(\ell)} \gets  \boldsymbol\Sigma^{\ast}$
   \Else
        \State $\textbf{V}^{(\ell)}_k \gets \textbf{V}_k^{(\ell-1)}$
   \EndIf
  \EndFor 
  \State $\boldsymbol\tau^\ast \sim q(\boldsymbol\tau^\ast|\boldsymbol\tau^{(\ell-1)})$ \Comment{$\boldsymbol\tau^\ast$ proposal made outside $k=1,\ldots,L$ loop; see text}
  \State $\Sigma^\ast \gets \Sigma(\lambda_1^{(\ell)},\theta_1^{(\ell)},\textbf{V}_1^{(\ell)}, \tau_1^\ast, \dots, \lambda_k^{(\ell)},\theta_k^{(\ell)},\textbf{V}^{(\ell)}_k, \tau_k^\ast, \dots, \tau_{L-1}^\ast,  \lambda_L^{(\ell)},\theta_L^{(\ell)},\textbf{V}_L^{(\ell)})$
  \State $\alpha \gets |{\boldsymbol\Sigma^\ast}^{-1}\boldsymbol\Sigma^{(\ell)}|^{1/2} \exp\left\{  -\frac{1}{2} \mathbf{Z}^T\left({\boldsymbol\Sigma^{\ast}}^{-1}\mathbf{Z} - {\boldsymbol\Sigma^{(\ell)}}^{-1}\mathbf{Z}\right) \right\}\dfrac{q(\boldsymbol\tau^{(\ell-1)}|\boldsymbol\tau^\ast)}{q(\boldsymbol\tau^\ast|\boldsymbol\tau^{(\ell-1)})}$
  \If{$\text{U} \leq \min(1,\alpha)$}
        \State $\boldsymbol\tau^{(\ell)} \gets \boldsymbol\tau^\ast$ and $\boldsymbol\Sigma^{(\ell)} \gets  \boldsymbol\Sigma^{\ast}$
   \Else
        \State $\boldsymbol\tau^{(\ell)} \gets \boldsymbol\tau^{(\ell-1)}$
   \EndIf
\EndFor
\end{algorithmic}
\end{algorithm}


The proposal for the set of change points $\boldsymbol\tau$ can be difficult to postulate, even in the case where the number of change points $L$ is known and fixed. One strategy is to update a single element of $\boldsymbol\tau$ at each Gibbs iteration, using symmetric random walk proposals for a randomly selected $k$-th element. Assume without loss of generality that $t_j = t_{j-1}+1,$ for $j=2,\ldots,m.$ Then the symmetric proposal is $\tau_k^\ast = \tau_k^{(\ell-1)}+1$ or $\tau_k^\ast = \tau_k^{(\ell-1)}-1,$ each with probability 1/2. However, at the boundaries $\tau_1 = 0$ or $\tau_{L-1},$ or adjacent to them, or when there are adjacent $\tau_{k+1} = \tau_k +1$ or $\tau_{k+1} = \tau_k + 2$ (or similarly for $\tau_{k-1} = \tau_k-1,$ etc.), it is necessary to perform a Hastings correction  \citep[analogous to the case in][]{carlin1992hierarchical}. 
Explicitly, the boundary corrections for unit steps after proposing to update the $k$-th change point are \[\frac{q\left(\tau_k^{(\ell-1)}\mid\tau^\ast_k\right)}{q\left(\tau^\ast_k \mid \tau_k^{(\ell-1)} \right)} = \begin{cases}0 & \text{if } \{\tau_{k-1}^{(\ell-1)} = \tau_{k}^{(\ell-1)}-1\} \land \{\tau_{k+1}^{(\ell-1)} = \tau_{k}^{(\ell-1)}+1\}; \\
1 & \text{if } \{\tau_{k-1}^{(\ell-1)} = \tau_{k}^{(\ell-1)}-1\} \land \{\tau_{k+1}^{(\ell-1)} = \tau_{k}^{(\ell-1)}+2\} \text{ OR } \\
  & \hphantom{\text{if }} \{\tau_{k-1}^{(\ell-1)} = \tau_{k}^{(\ell-1)}-2\} \land \{\tau_{k+1}^{(\ell-1)} = \tau_{k}^{(\ell-1)}+1\}, \\
  & \text{toward only allowed movement, i.e.,} \tau^\ast_k=\tau_{k}^{(\ell-1)}+1 \text{ OR } \tau^\ast_k=\tau_{k}^{(\ell-1)}-1, \text{respectively;} \\
1/2 & \text{if } \{\tau_{k-1}^{(\ell-1)} = \tau_{k}^{(\ell-1)}-1\} \land \{\tau_{k+1}^{(\ell-1)} \geq \tau_{k}^{(\ell-1)}+3\} \text{ OR }  \\
  & \hphantom{\text{if }}\{\tau_{k-1}^{(\ell-1)} \leq \tau_{k}^{(\ell-1)}-3\} \land \{\tau_{k+1}^{(\ell-1)} = \tau_{k}^{(\ell-1)}+1\}, \\
  & \text{toward only allowed movement, i.e.,} \tau^\ast_k=\tau_{k}^{(\ell-1)}+1 \text{ OR } \tau^\ast_k=\tau_{k}^{(\ell-1)}-1, \text{respectively;} \\
2 & \text{if } \{\tau_{k-1}^{(\ell-1)} = \tau_{k}^{(\ell-1)}-2\} \land \{\tau_{k+1}^{(\ell-1)} = \tau_{k}^{(\ell-1)}+2\}, \\
  & \text{toward either allowed movement, i.e.,} \tau^\ast_k=\tau_{k}^{(\ell-1)}+1 \text{ OR } \tau^\ast_k=\tau_{k}^{(\ell-1)}-1; \\
2 & \text{if } \{\tau_{k-1}^{(\ell-1)} = \tau_{k}^{(\ell-1)}-2\} \land \{\tau_{k+1}^{(\ell-1)} \geq \tau_{k}^{(\ell-1)}+3\} \text{ OR } \\
  & \hphantom{\text{if }}\{\tau_{k-1}^{(\ell-1)} \leq \tau_{k}^{(\ell-1)}-3\} \land \{\tau_{k+1}^{(\ell-1)} = \tau_{k}^{(\ell-1)}+2\}, \\
  & \text{toward movements } \tau^\ast_k=\tau_{k}^{(\ell-1)}+1 \text{ OR } \tau^\ast_k=\tau_{k}^{(\ell-1)}-1, \text{respectively;} \\
1 & \text{if } \{\tau_{k-1}^{(\ell-1)} = \tau_{k}^{(\ell-1)}-2\} \land \{\tau_{k+1}^{(\ell-1)} \geq \tau_{k}^{(\ell-1)}+3\} \text{ OR } \\
  & \hphantom{\text{if }}\{\tau_{k-1}^{(\ell-1)} \leq \tau_{k}^{(\ell-1)}-3\} \land \{\tau_{k+1}^{(\ell-1)} = \tau_{k}^{(\ell-1)}+2\}, \\
  & \text{toward movements } \tau^\ast_k=\tau_{k}^{(\ell-1)}-1 \text{ OR } \tau^\ast_k=\tau_{k}^{(\ell-1)}+1, \text{respectively;} \\
1 & \text{otherwise}.\end{cases}\] 

Observe that for the case \(L=2,\) we may use \[\phi_1(t) = \mathbf{1}\{t \leq \tau\}, \quad \phi_2(t) = \mathbf{1}\{t > \tau\},\] and the prior \[\tau \sim U\{t_1,t_2,\ldots,t_m\}.\] In this configuration, the posterior of \(\tau | \text{others}\) is a discrete random variable with mass \[\mathbb{P}(\tau = \tau_j | \text{others}) = \left(1 + \displaystyle\sum_{\ell \neq j}|\boldsymbol\Sigma^{-1}_{\tau_\ell}\boldsymbol\Sigma_{\tau_j}|^{1/2} \exp\left\{ -\frac{1}{2} \mathbf{Z}^T\left(\boldsymbol\Sigma_{\tau_\ell}^{-1}\mathbf{Z} - \boldsymbol\Sigma_{\tau_j}^{-1}\mathbf{Z}\right) \right\}\right)^{-1},\] and therefore it is possible to sample exactly from $\tau$ without a Metropolis step, unlike in Algorithm \ref{alg:gibbs}. This case has much faster mixing times for the MCMC, in general, but the numerical evaluation of $m$ inverse matrices can lead to slower computing time of each iteration, and thus we expect better temporal scalability of the random-walk approach. For the simulations and case study presented in this work, when $L=2$ we used the exact sampler, and the random walk sampler when $L \geq 3.$

\subsection{Determining the number \textit{L} of advection direction changes}

Model selection can be done via information criteria, but the spatial dependence prevents usage of AIC or BIC methods, as discussed, for example, in \citet{hoeting2006model}. A fully Bayesian method such as WAIC \citep{watanabe2010asymptotic,gelman2014understanding} could be an alternative approach, but its implementation based on the MCMC approach of Algorithm \ref{alg:gibbs} would require conditional simulations. 
Therefore, we use \citet{spiegelhalter2002bayesian}'s DIC, as its implementation is straightforward \citep[see also][]{celeux2006deviance}. Let \[\overline{\log \mathcal{L}(\mathbf{z} \mid \boldsymbol\zeta)} = \frac{1}{M} \sum_{\ell = 1}^M \log( \mathcal{L}(\mathbf{z} \mid \boldsymbol\zeta_\ell)),\] where $\boldsymbol\zeta_\ell$ are the sampled posterior parameters from Algorithm \ref{alg:gibbs}, and \begin{equation}\text{DIC} = - 4 \overline{\log \mathcal{L}(\mathbf{z} \mid \boldsymbol\zeta)} + 2 \log \mathcal{L}(\mathbf{Z}|\bar{\boldsymbol\zeta}),\label{eq:dic}\end{equation} 
where $\bar{\boldsymbol\zeta}$ is the average posterior sample of the parameters. As defined, a smaller DIC is preferred when selecting a model for $L.$

\section{Simulation study}\label{sec4sim}

We will perform a brief simulation study. We considered:

\begin{itemize}
\item
  A set \(\mathcal{S}\) of \(n=36\) spatial locations, either on a \emph{regular} grid or a \emph{random} configuration fixed across replications. We sample a Gaussian process at \(m=12+1\) equidistant time points \(\mathcal{T} \cup \mathcal{T}_{1+}= \{1, 2, \ldots, 12\} \cup \{13\}.\) 
\item
  The data was simulated using \(\tau = 6;\) as well as
  \(\mathbf{V}_1 = \begin{pmatrix}0.25 & 0.25\end{pmatrix}^T,\)
  \(\mathbf{V}_2 = \begin{pmatrix}0.00 & -0.25\end{pmatrix}^T.\)
\item
  The spatial dependence parameters are \(\theta_1 = 0.1,\) \(\theta_2 =0.2.\) These values correspond to a \textit{practical range} of 0.3 and 0.6 distance units in \(\mathcal{S},\) respectively. We use \(\lambda_1 = \lambda_2 = 1\) for spatial variances.
\item
  The likelihood function is built using only data observed in \(\mathcal{S} \times \mathcal{T};\) we will use the data in \(\mathcal{S} \times \mathcal{T}_{1+}\) to assess prediction (1-unit of time future forecast).
\item 
  We consider our dynamic advection model using either $L=2$ (correct) or $L=3$ (overparametrized). We also consider a Bayesian formulation of the Gupta-Waymire model, corresponding to the particular case $L=1,$ using the same priors and no change function \(\phi.\) The Bayesian Gupta-Waymire model is also fitted via MCMC.
\item
  We repeated each scenario \(B=10\) independent times. MCMC was run for 1000 steps, with a burn-in period of 200. This takes roughly 5 minutes per simulation when $L=2,$ or 15 minutes per simulation when $L=3.$ 
\end{itemize}


In addition to evaluated DIC, all fitted models will be tested using a one-step-ahead forecast, using posterior estimates of $\hat{y}(\mathbf{s}, t= 13).$ We will consider the posterior mean squared prediction error relative to the true values of $y(\mathbf{s}, t = 13)$ generated during the simulation step. The prediction error (PE) will be calculated as \[\text{PE}(\text{model}) = \sqrt{(Bn)^{-1}\sum_{b=1}^B \sum_{i=1}^n[\hat{y}_{b, \text{model}}(\mathbf{s}_i, t= 13) - y_b(\mathbf{s}_i, t = 13)]^2},\] where $b= 1,\ldots,B$ represent the independent simulations ($B=10$ in this study).

\begin{figure}[t]
\centering
\includegraphics[width = 0.95\textwidth]{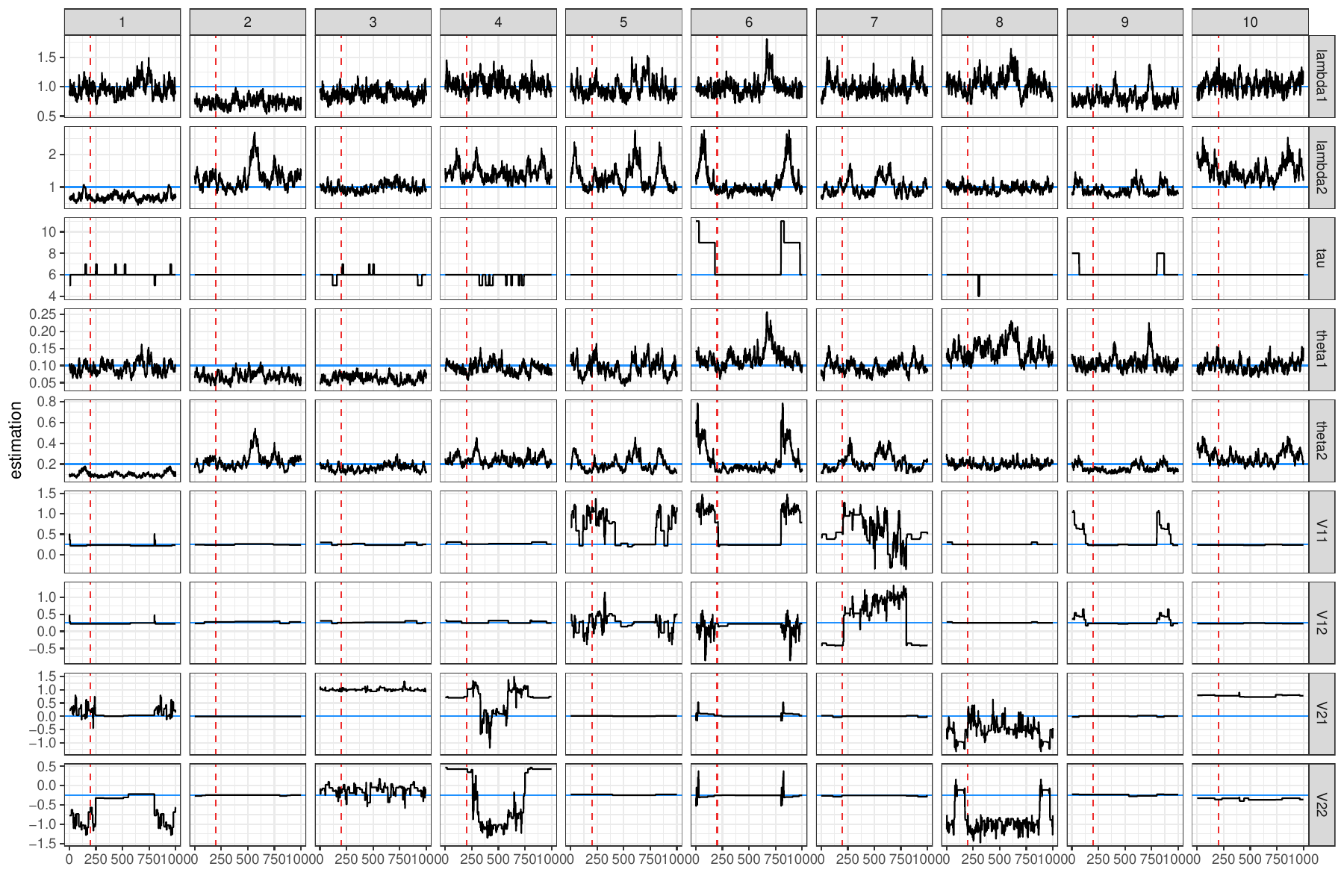} 
\caption{Trace plots across 10 independent MCMC simulations (columns) versus posterior samples of the parameters (rows). The vertical red line indicates the end of the burn-in period. In this simulation study, the spatial locations are on random sites (uniformly distributed), but fixed across simulations.}
\label{fig:cp=TRUE-grid=aleatorio-plot=chain}
\end{figure}

The results of the simulated study are shown for $L=2$ in Figure \ref{fig:cp=TRUE-grid=aleatorio-plot=chain}. We will report only the results of the random grid, as the results from the regular grid are similar. We obtain stationarity of \(\boldsymbol\lambda\) posterior samples without issues, centered on the correct (that is, simulated) values. The parameters \(\boldsymbol\theta\) are sensitive to choice of priors. In this simulation, we used an Exponential(6), so the expected priors are informative, not too far from the ``true'' values from which data was simulated from (\(\theta_1 = 0.1, \theta_2 = 0.2\)). The spatial dependence parameter in spatial covariance models is known to be difficult to estimate and require informative priors; see the discussion in \citet{zhang2004inconsistent}. The Metropolis steps for \(\mathbf{V}\) tend to almost always reject proposals (we are using random-walk normal proposals). We have thus made very small proposal steps for them; for the case study, larger MCMC chains will be used get reliable posterior samples of \(\mathbf{V}_k.\) Lastly, the estimate of $\tau$ is concentrated on the ``true'' simulated value of $\tau=6,$ but since for $L=2$ the posterior distribution of $\tau$ is conjugated, and therefore we can conclude that the posterior mass is indeed concentrated on the change point value. The case when $L=3$ also showed poor mixing of $\tau_1, \tau_2.$

We remark that the Gupta-Waymore model suffers from the same issues as the Dynamic Advection model when posterior sampling \(\theta\) and \(\mathbf{V}.\) Prediction is done by averaging ordinary Kriging obtained from the posterior sample of the parameters. The average squared prediction error (PE), and its standard deviation, for the models are shown in Table \ref{tab:EPsimData}. Forecast is performed one hour into the future.  Moreover, Table \ref{tab:EPsimData} also shows the average Deviance Information Criteria (DIC) defined in Equation \eqref{eq:dic}, as well as its observed standard deviation across simulations. DIC was evaluated after discarding burnin period and at every 10 samples. 
Both PE and DIC criteria were satisfactory in identifying the correct $L,$ on average; but had large standard deviations across independent simulations. The DIC was systematically  larger (across independent simulations) for the Gupta-Waymire (incorrect) model. However, DIC struggled to differentiate the $L=2$ (correct) from the $L=3$ (overparametrized) cases. In those situations, the PE criteria performed better.

\begin{table}
\caption{Average squared prediction error (PE) and DIC, as well as their standard deviation, for the Dynamic Advection ($L=2,3$ change points), and Gupta-Waymire models, considering spatial sample sites both in a regular grid and random (fixed across simulations).}
\label{tab:EPsimData}
\centering
\begin{tabular}{l|cc|cc|cc}
\multirow{2}{*}{Grid}  & \multicolumn{2}{c|}{Bayesian Gupta-Waymire ($L=1$)} & \multicolumn{2}{c|}{Dynamic Advection ($L=2$)} & \multicolumn{2}{c}{Dynamic Advection ($L=3$)} \\
        & PE (s.d) & DIC (s.d) & PE (s.d) & DIC (s.d) & PE (s.d) & DIC (s.d) \\
\hline
Regular & 1.09 (0.26) & 585.5 (47.3)\hphantom{0} & 0.89 (0.33) & 201.61 (84.43) & 1.05  (0.25) & 220.45 (75.90) \\
Random  & 1.28 (0.65) & 625.7 (140.0) & 0.67 (0.26) & 149.50 (56.23) & 1.05 (0.32) & 156.76 (76.04) \\
\end{tabular}
\end{table}

\section{Case study: extreme weather event in Southeastern Brazil}

\begin{figure}[htb]
\centering
\includegraphics[width=0.35\linewidth]{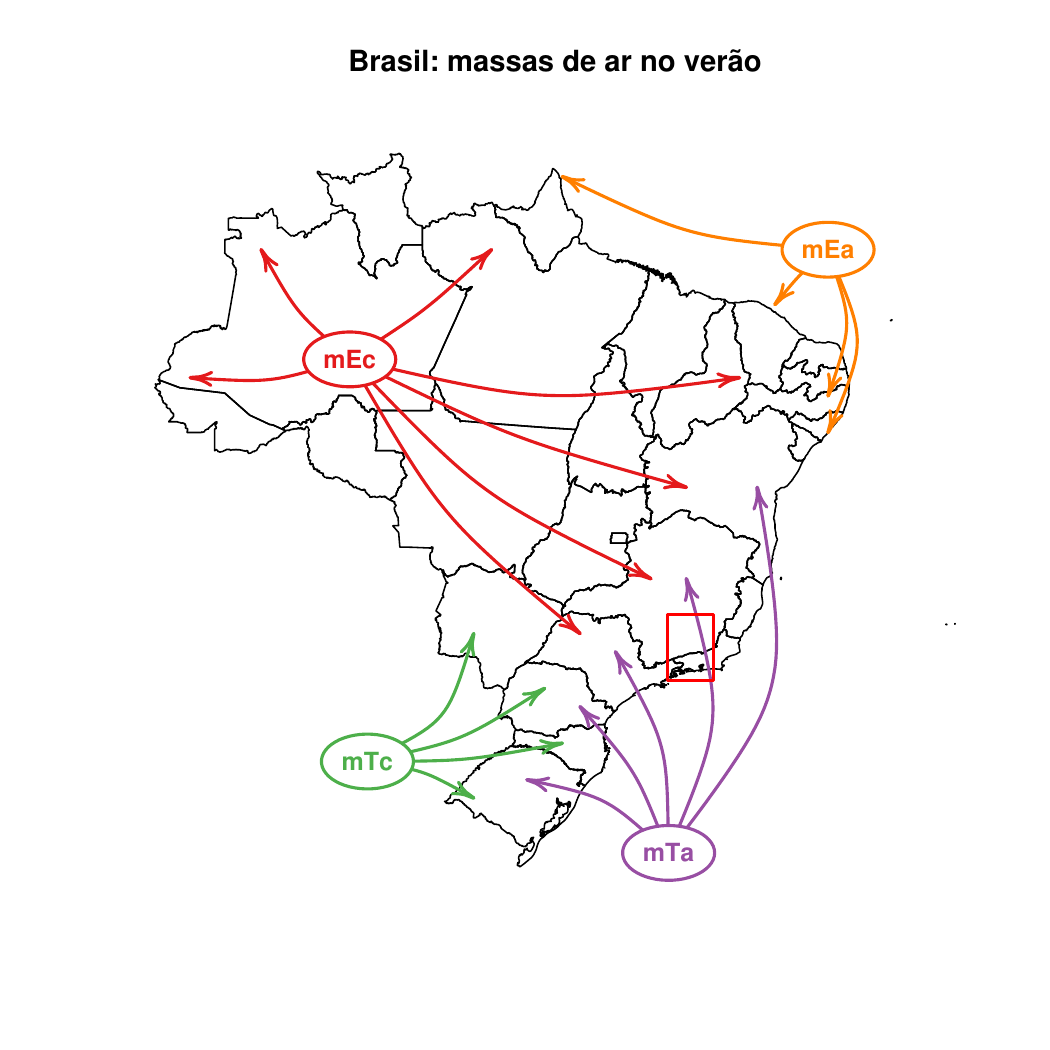} 
\includegraphics[width=0.35\linewidth]{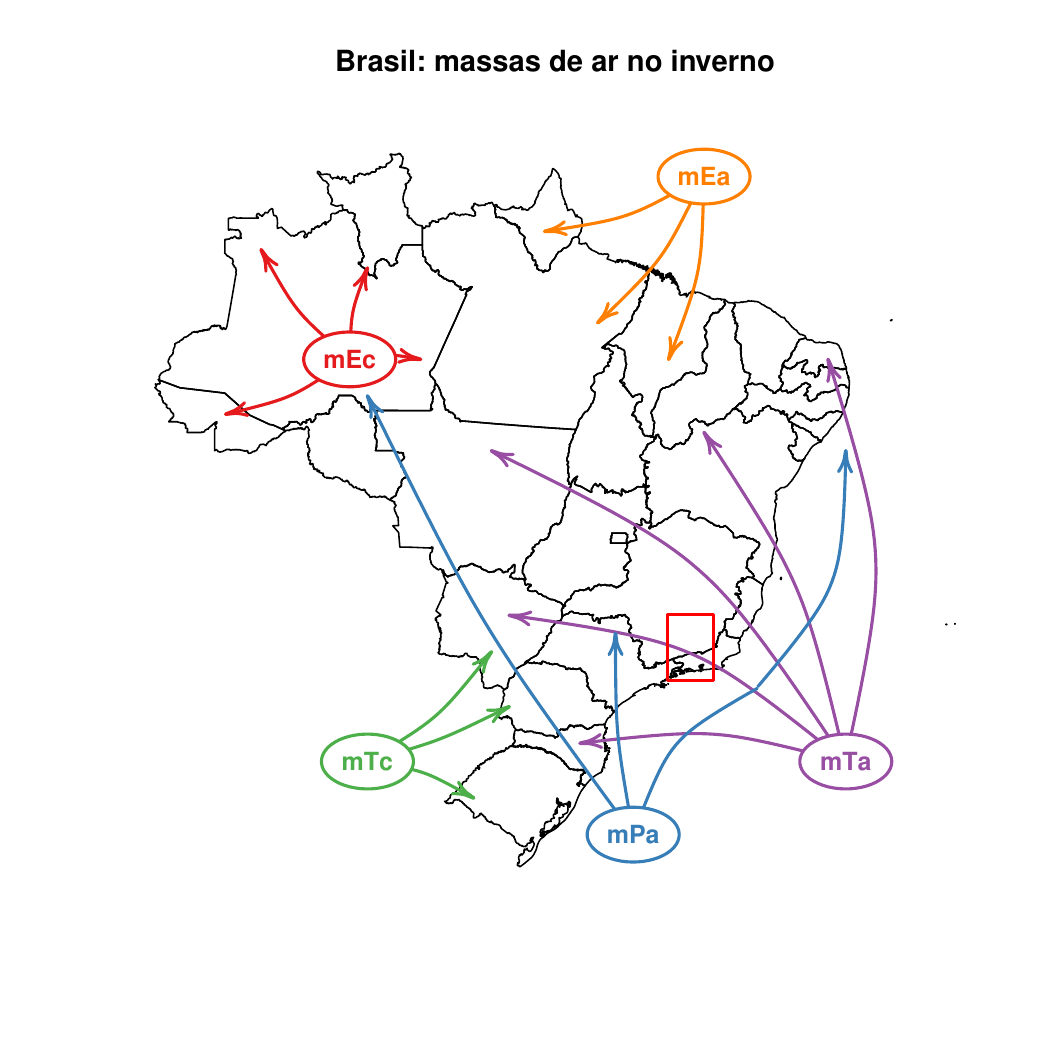} 
\includegraphics[width=0.20\linewidth]{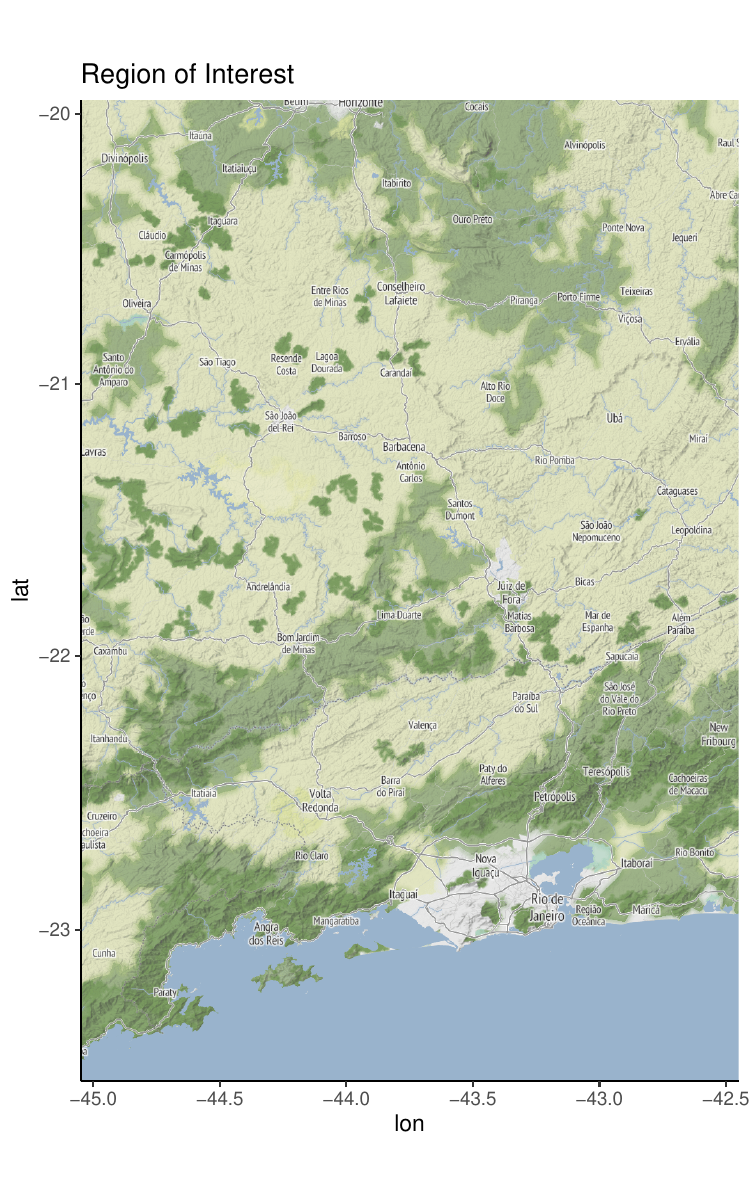} 
\caption{South American air masses during summer (left) and winter (middle); region of interest (right).}
\label{fig:mass}
\end{figure}

\begin{figure}[ht]
\centering
\includegraphics[width = 0.9\textwidth]{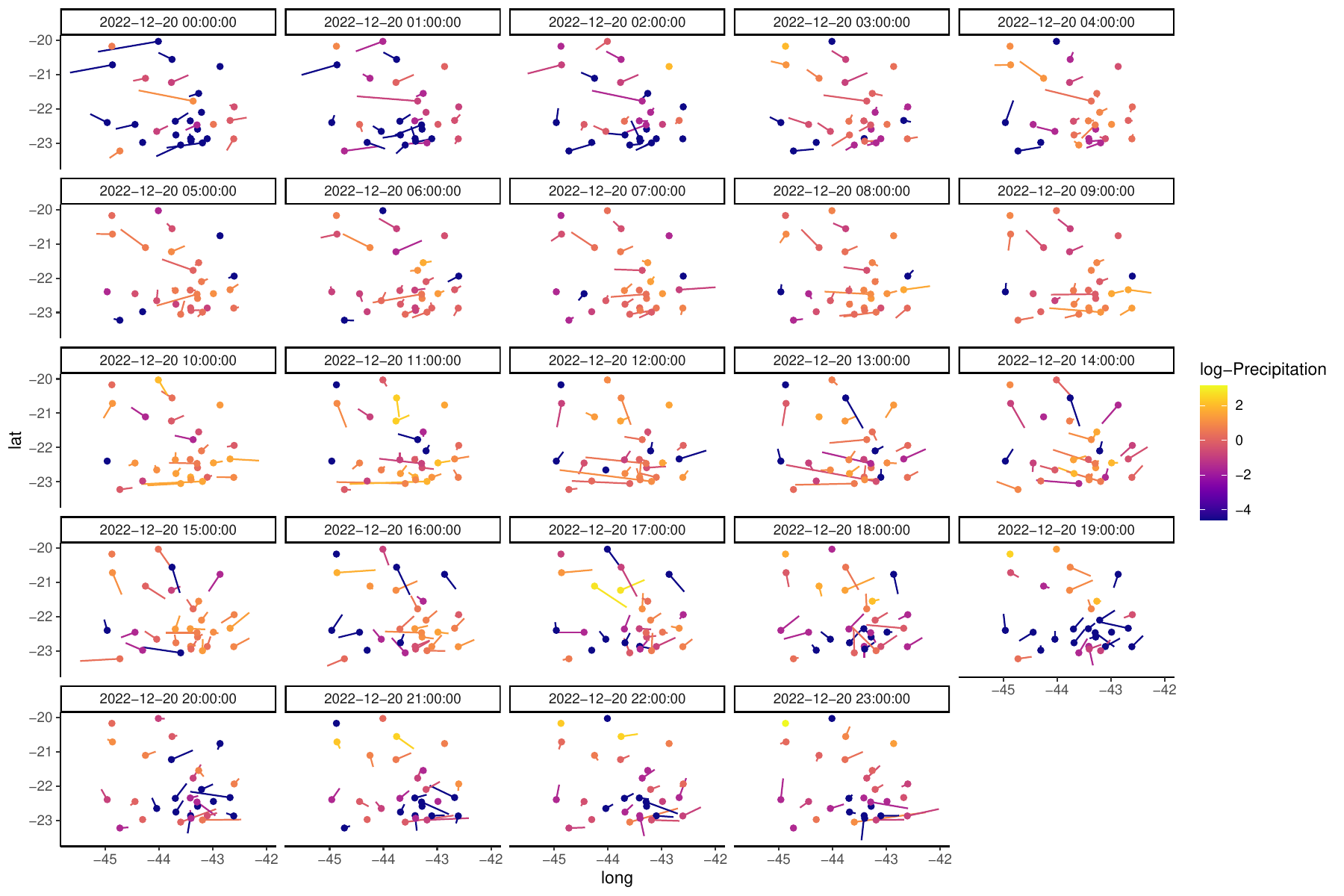} 
\caption{Hourly aggregated data from rainfall stations, where log-rainfall intensity (color) and wind direction (segments, rotated wind flow), in the region of interest (Rio de Janeiro and southern Minas Gerais states), December 20th, 2022.}
\label{fig:dados_reais_points}
\end{figure}

The spatiotemporal data used in this study were obtained from the INMET - National Institute for Meteorology, an agency of the Ministry of Agriculture of Brazil. Rainfall weather stations data, such as temperature, humidity, atmospheric pressure, precipitation, wind direction and speed, and solar radiation, are collected automatically and reported in INMET's website, updated every 15 minutes: \url{https://mapas.inmet.gov.br/}. The southeastern region of Brazil has shocks between Equatorial continental winds and tropical Atlantic winds during summer, both of which are warm and humid (Figure \ref{fig:mass}, left). The interaction between these air masses causes frequent, intense tropical rainstorms in the region of interest, whose map is shown in Figure \ref{fig:mass}, right. We used 24 hours of data from the day of 2022-12-20, as the region of interest had an extreme weather event on the day, with large volume of rainfall \footnote{News report about the weather event is available in \url{https://g1.globo.com/rj/regiao-dos-lagos/noticia/2022/12/20/chuva-atinge-todo-o-estado-do-rio-de-janeiro-praias-estao-com-alerta-de-ressaca.ghtml}, retrieved in September 2026.}. We aggregated data every 15 minutes to one hour blocks, to reduce variation. Figure \ref{fig:dados_reais_points} shows the spatial sites, with each panel displaying one hour of aggregated data. The resulting wind direction is shown with line segments, with length proportional to wind speed and direction indicating the observed flow direction (and not the direction where the wind is coming from). The color intensity corresponds to the registered log-Precipitation volume; to be precise, $\log(\text{precipitation}+0.01).$.

To decide which model best fit the data, we used the subsequent hour of rainfall (2022-12-20 from midnight to 1am), comparing mean squared error of forecast \textit{a posteriori} for one time period, using Gupta-Waymire's model with MCMC; and the dynamic advection model proposed in this work (DA) with different choices of $L=2,3,4.$ We also compared the DIC criteria. We also included an additional parameter $\lambda_0,$ which served as a nugget effect, and smoothed out the Kriging maps, in all model formulations. The results are summarized in Table \ref{tab:EPrealData}. We can readily discard Gupta-Waymire's model, having worse PE and DIC. Among the dynamic advection models, even though the case $L=3$ has lower DIC and PE than $L=2,$ we decided to present the results for $L=2,$ as the improvement is only small (3\% worse PE, 5\% worse DIC). We observe that $L=4$ is starting to deteriorate, having worse performance than $L=3,$ but still has very close numerically.

\begin{table}[htb]
\caption{Squared average prediction error (standard deviation) for the case study data, using one hour into the future of the model calibration dataset, and DIC.}
\label{tab:EPrealData}
\centering
\begin{tabular}{l|rr}
Model & PE (s.d.) & DIC \\
\hline
Bayesian Gupta-Waymire ($L=1$) & 11.11 (13.13) & 1926.6 \\
Dynamic Advection ($L=2$) & 9.30 (12.88) & 1454.1 \\ 
Dynamic Advection ($L=3$) & 9.03 (12.26) & 1384.6 \\
Dynamic Advection ($L=4$) & 9.11 (12.29) & 1399.7 \\
\end{tabular}
\end{table}

\begin{figure}[ht]
\centering
\includegraphics[width = 0.9\textwidth]{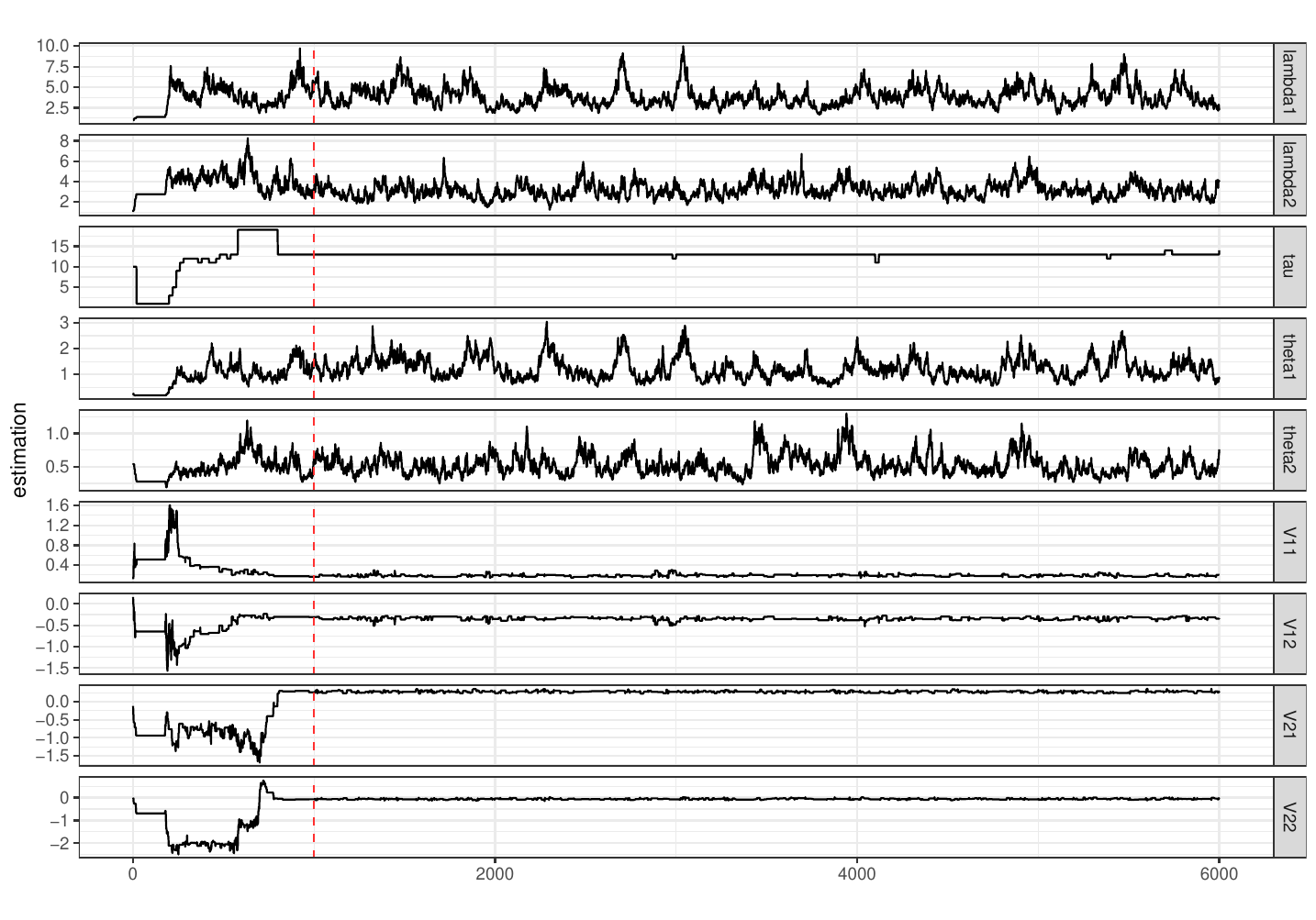} 
\caption{Traceplot of the posterior distribution of $\lambda_0\mid\text{others},$ $\lambda_1\mid\text{others}$ and $\lambda_2\mid\text{others},$ $\theta_1\mid\text{others},$ $\theta_2\mid\text{others},$ $\tau\mid\text{others}$ as well as $\mathbf{V}_1\mid\text{others}$ and $\mathbf{V}_2\mid\text{others},$ for the ROI in Rio de Janeiro and south of Minas Gerais.}
\label{fig:realdata_cp=TRUE_chainplot}
\end{figure}

\begin{figure}[ht]
\centering
\includegraphics[width = 0.9\textwidth]{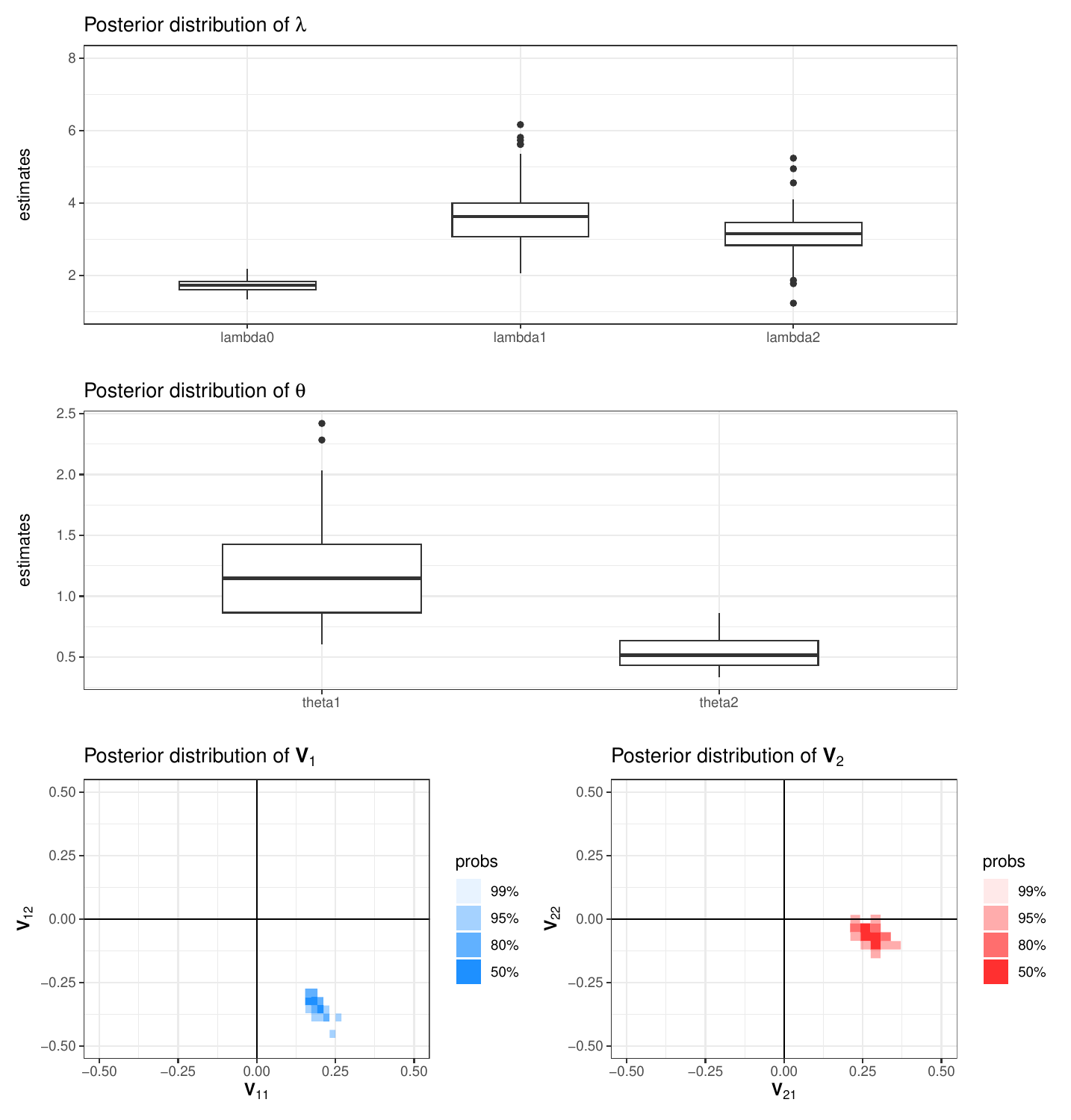} 
\caption{Posterior distribution of $\lambda_0\mid\text{others},\lambda_1\mid\text{others}$ and $\lambda_2\mid\text{others}$ (spatial variances, top row), $\theta_1\mid\text{others},\theta_2\mid\text{others}$ (spatial dependence, mid row) as well as $\mathbf{V}_1\mid\text{others}$ and $\mathbf{V}_2\mid\text{others}$ (advection direction; spatial histograms in $\mathbb{R}^2,$ bottom row). The posterior mode of $\tau \mid \text{others}$ is 13:00.}
\label{fig:realdata_cp=TRUE_posterior_parameter}
\end{figure}

\begin{figure}[ht]
\centering
\includegraphics[width = \textwidth]{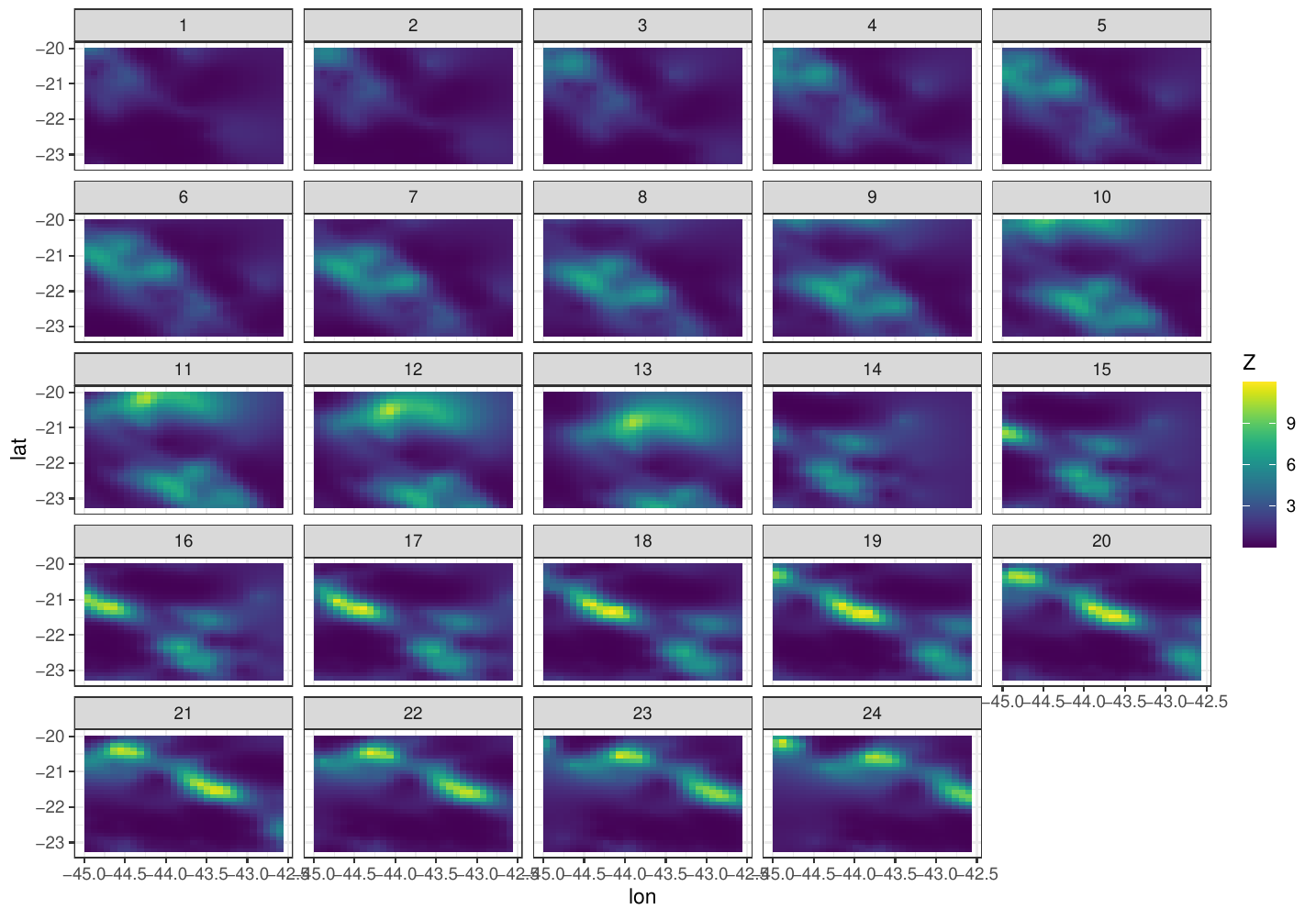} 
\caption{Predicted maps with posterior average of $Z \mid \{Z_{\text{obs}}, \text{others}\}.$ The posterior mode of $\tau \mid \text{others}$ is 13:00.}
\label{fig:realdataMapExp}
\end{figure}

The fitted model with $L=2$ detected a change of wind direction at posterior mode $\hat{\tau}\mid\text{others} = \text{13:00}.$ In Figure \ref{fig:realdata_cp=TRUE_chainplot}, the posterior samples of the parameters are shown in a traceplot. We observe that the parameters have attained stationarity of their respective posterior distributions; however $\tau$ does not move much after the burn-in period. Since for the case $L=2$ the posterior is conjugated, we can conclude it is merely very concentrated in $\tau = \text{13:00}.$ After discarding the first 1000 observations as burn-in, Figure \ref{fig:realdata_cp=TRUE_posterior_parameter} shows the posterior distribution of $\lambda_0\mid\text{others},\lambda_1\mid\text{others}$ and $\lambda_2\mid\text{others}$ (spatial variances, top row), $\theta_1\mid\text{others},\theta_2\mid\text{others}$ (spatial dependence, mid row) as well as $\mathbf{V}_1\mid\text{others}$ and $\mathbf{V}_2\mid\text{others}$ (advection direction; spatial histograms in $\mathbb{R}^2,$ bottom row). Observe that the posterior nugget effect $\lambda_0\mid\text{others}$ is comparatively small with respect to the posterior spatial variances; that the spatial dependence $\theta_1\mid\text{others}$ is stronger than $\theta_2\mid\text{others},$ which suggests a stronger SE wind effect prior to $\tau = \text{13:00},$ followed by a smaller effect toward E after $\tau = \text{13:00}.$ In Figure \ref{fig:realdataMapExp}, we show the predicted posterior maps of $Z,$ by exponential transforming the Kriging maps for the log-precipitation data. We observe a drift south-southest in the first hours of the day, followed by a drift east after 13:00, as predicted from our model. The model suggests that Equatorial winds may have dominated cloud movement on that day.

\section{Discussion}

The dynamic advection (AD) models proposed here have superior performance compared to the Gupta-Waymire and Cox-Isham models on both regular and irregular grids. This indicates that, for phenomena involving dynamic advection, relying on rigid processes, such as the Gupta-Waymire model, which assumes a single advection direction, results in a loss of predictive accuracy.

The ability to estimate the change point confirms dynamic effects on precipitation during Brazil's summer. Using precipitation data from southeastern Minas Gerais and Rio de Janeiro region showed the effect of shocks of different air masses, which may cause extreme weather events. This also aligns with the finding that change-point estimation in AD models improved prediction accuracy in the simulation studies.

Furthermore, the Bayesian approach renders the distinction between ``frozen'' and ``non-frozen'' fields, discussed in \citet{cox1988spatial}, unnecessary. By performing Bayesian Kriging using the posterior distribution of the parameters, we observe the diffusion of the advection effect proposed by \citet{cox1988spatial}, which is more flexible than the rigid transport mechanism of \citet{gupta1987taylor}. The Bayesian methodology allows for adaptive advection; conversely, minimal variation in the posterior parameters, particularly $\mathbf{V}_k,$ results in more rigid advection within the spatio-temporal Kriging framework. 

However, it is important to note that estimating the dynamic advection models proposed here entails significant computational cost, inherent to the complexity of the covariance structure and the Bayesian inference via MCMC. This computational time may pose a practical limitation for applying the method to higher-dimensional spatiotemporal datasets or in scenarios requiring faster analyses, potentially necessitating the use of high-performance computing resources. SPDE-based approaches for models with advection have been discussed in \citet{carrizo2022general}, which may allow the use of INLA-based approximations \citep{rue2009approximate}. This is a topic of future work.

Another limitation is the use of log-Gaussian processes, particularly in the sense of large periods / regions with no rainfall. Regions with no rainfall are impossible to appear if the data was indeed log-Gaussian. The authors are currently exploring a fusion of dynamic advection models with rolling windows of spatial disks and the temporal Bartlett-Lewis rectangular process, such as discussed in \citet{cox1998stochastic,northrop1998clustered,northrop2024stochastic}. This is also a topic of future work.


\section*{Acknowledgments}
Pedro N. Ramos received support from FAEPEX/PRP Unicamp Foundation process 2536/20. Guilherme Ludwig acknowledges funding from FAPESP processes 2019/03517-0, 2023/02538-0, 2024/05242-7, and 2026/07255-4.



\subsection*{Financial disclosure}

None reported.

\subsection*{Conflict of interest}

The authors declare no potential conflict of interests.

\bibliographystyle{apalike}  
\bibliography{references}

\end{document}